\documentstyle[12pt,epsf]{article}

\newcommand{\del}{\partial}
\newcommand{\Gam}{\Gamma}
\newcommand{\dis}{\displaystyle}
\newcommand{\ba}{\begin{eqnarray}}
\newcommand{\ea}{\end{eqnarray}}
\newcommand{\non}{\nonumber\\}
\newcommand{\nom}{\nonumber}
\newcommand{\vpi}{\varpi}
\newcommand{\tpi}{\tilde{\varpi}}
\newcommand{\kae}{\mbox{K\"ahler}}
\newcommand{\kei}{\mbox{\Large\mbox{$\kappa$}}}

\newcommand{\Mfunction}[1]{#1}
\newcommand{\re}{\mbox{Re}}
\newcommand{\im}{\mbox{Im}}
\begin{document}

\begin{titlepage}
\nopagebreak
\begin{flushright}
March 2000\hfill 
hep-th/0003165
\end{flushright}

\vfill
\begin{center}
{\LARGE K\"ahler Potential of Moduli Space }

~

{\LARGE   in Large Radius Region of Calabi-Yau Manifold }

\vskip 12mm

{\large Katsuyuki~Sugiyama${}^{\dag}$}

\vskip 10mm
{\sl Department of Fundamental Sciences}\\
{\sl Faculty of Integrated Human Studies, Kyoto University}\\
{\sl Yoshida-Nihon-Matsu cho, Sakyo-ku, Kyoto 606-8501, Japan}\\
\vspace{1cm}

     \begin{large} ABSTRACT \end{large}
\par
\end{center}
\begin{quote}
 \begin{normalsize}

We study a {\kae} potential $K$ in the large radius region 
of a Calabi-Yau $d$-fold $M$ embedded in $CP^{d+1}$.
It has a {\kae} parameter t that describes a 
deformation of the A-model moduli.
Also the metric, curvature and hermitian two-point functions 
in the large volume region
are analyzed. 
We use a result of our previous paper in the B-model of the mirror. 
We perform an analytic continuation of a parameter to the 
large complex structure region. By translating the result in the A-model 
  side of $M$, we determine the $K$.
The method is not restricted to this specific model
and we  apply the recipe to complete intersection Calabi-Yau cases.

\end{normalsize}
\end{quote}

\mbox{}\hspace{21mm}
\vfill 
\noindent
\rule{7cm}{0.5pt}\par
\vskip 1mm
{\small \noindent ${}^{\dag}$  
E-mail :\tt{} sugiyama@phys.h.kyoto-u.ac.jp}
\end{titlepage}
\vfill

\section{Introduction}


D-branes play important roles to describe the solitonic modes in string 
theory and could make clear dynamics in strong coupling regions. The 
physical observables of D-brane's effective theories have dependences on 
moduli of compactified strings or wrapped D-branes. 
We expect that properties of compactified internal spaces essentially
control 
non-perturbative effects in low energy theories. 
In this paper, we focus on the type II superstring compactified on
Calabi-Yau manifold and study its topological sector
from the point of view of 
topological sigma models (A- and B-models)\cite{W} 
to investigate properties of moduli spaces.
Because both the A- and B-models are topological theories, they are
characterized by their two-point and three point functions that 
play important roles as the constituent blocks 
in these models\cite{GP}-\cite{KS7}.
Topological metrics are two-point functions and receive no quantum 
corrections. On the other hand, 
the three-point functions of the A-model have information about the
fusion structure of observables. 
The remaining fundamental blocks are a {\kae} potential $K$ and 
associated hermitian two-point functions. 
They are hermitian and describe correlations of topological ant
anti-topological sectors\cite{tt},\cite{CFT},\cite{KS5},\cite{KS7}. 
Also the $K$ has information about
intersections of homology cycles with even dimensions in the A-model
case.

The aim of this paper is to develop a concrete method to 
construct {\kae} potentials 
applicable in the large radius regions of Calabi-Yau $d$-folds and to
investigate their properties in order to understand structures of the
moduli spaces.
We present formulae of {\kae} potentials 
for $d$ dimensional Calabi-Yau manifolds 
explicitly.

The paper is organized as follows. 
In section 2, we explain a mirror manifold paired with a Calabi-Yau 
$d$-fold embedded in $CP^{d+1}$. We also explain the results in \cite{KS7}
about a {\kae} potential $K$ in the small complex structure region of 
the B-model in order to fix notations.
In section 3, we introduce a set of periods valid in the 
large complex structure region. By relating two sets of periods in the 
large and small complex structure regions, 
we construct a formula of the $K$ applicable in the 
large complex structure region of the B-model. In sections 4 and 5, we
construct a mirror map and a {\kae} potential. The scalar
curvature of the {\kae} moduli is investigated.
The set of correlation functions associated with the {\kae} moduli are
calculated  in the large radius region of the A-model. 
A concrete application of our result is explained in the quintic case in 
 section 6. Also the result there is generalized to propose a formula of 
the {\kae} potential of a Calabi-Yau $d$-fold
in the complete intersection type in section 7. Section 8 is devoted to
conclusions and comments. In appendix A, we summarize several examples of
the expansion coefficients of a function $\hat{K}$ in lower dimensional
cases.

\section{Small Complex Structure Region}

We take a one-parameter family of Calabi-Yau $d$-fold $M$ realized as a
zero locus of a hypersurface embedded in a $CP^{d+1}$
\ba
M\,;\,p=X^N_1+X^N_2+\cdots +X^N_N-N\psi X_1X_2\cdots X_N=0\,.\nom
\ea
The $N$ is related with the complex dimension $d$ of M, $N=d+2$.
A mirror manifold $W$ paired with this $M$ is constructed as a orbifold 
divided by some maximally discrete group $G={\bf Z}_N^{\otimes (N-1)}$
\ba
W\,;\,\widehat{\{p=0\}/G}\,.\nom
\ea
When one thinks about Hodge structure of the $G$-invariant parts of the 
cohomology group $H^d(W)$, related Hodge numbers\cite{Dais} are written as
\ba
h^{d,0}=h^{d-1,1}=\cdots =h^{1,d-1}=h^{0,d}=1\,.\nom
\ea

In our previous paper\cite{KS7}, 
we study the formula of the {\kae} potential 
of the Calabi-Yau $d$-fold $W$
with a moduli parameter $\psi$ of the complex structure.

The $K$ is constructed by combining a set of periods $\tpi_k$ quadratically
\ba
&&e^{-K}=\sum_{k=1}^{N-1}I_{k}\tpi^\dagger_k\tpi_k\,,\label{KO}\\
&&I_k=\frac{1}{\pi^N\cdot N^{N+2}}(-1)^{k-1}
\left(\sin \frac{\pi k}{N}\right)^N\,,\non
&&\tpi_k(\psi)
=\left[\Gamma\!\left(\frac{k}{N}\right)\right]^N
\frac{(N\psi)^{k}}{\Gamma (k)}\non
&&\qquad \qquad \times\left[\sum_{n=0}^{\infty}
\left[\frac{\Gamma\!\left(\frac{k}{N}+n\right)}
{\Gamma\!\left(\frac{k}{N}\right)}\right]^N\frac{\Gamma (k)}{\Gamma (Nn+k)}
(N\psi)^{Nn}\right]\,.\nom
\ea
We determined the coefficients $I_k$ in the \cite{KS7}
by requiring  consistency conditions with the results of the CFT at 
the Gepner point.
This formula is valid in the small $\psi$ region because of the 
convergence of the series expansion. 

In the following section, we investigate the large complex structure 
region of the W. Its formula is important for the 
large radius analyses of the manifold $M$.

\section{Large Complex Structure Region}

We try to  rewrite the $K$ in a formula that is 
valid in the large $\psi$ by an 
analytic continuation.
First we choose a set of periods $\{\Omega_{m}\}$ ($m=0,1,\cdots ,N-2$) 
appropriate to describe the large complex 
structure region of the mirror W. 
A generating function of the $\Omega_m$ is defined 
by using a formal parameter $\rho$ with $\rho^{N-1}=0$
\ba
&&\sum^{N-2}_{m=0}\Omega_m\rho^m =\sqrt{\hat{K}(\rho)}
\cdot \varpi \left(\frac{\rho}{2\pi i};z\right)\,,\non
&&\vpi (v)=z^v\cdot \sum_{n=0}^{\infty}
\frac{a(n+v)}{a(v)}z^n\,,\,\,\,z=(N\psi)^{-N}\,,\label{series}\\
&&a(v)=\frac{\Gamma\!(Nv+1)}{[\Gamma\!(v+1)]^N}\,.\nom
\ea
Here we introduce a function $\hat{K}(\rho)$.
\ba
&&\hat{K}(\rho):=
\frac{\dis a\left(+\frac{\rho}{2\pi i}\right)}
{\dis a\left(-\frac{\rho}{2\pi i}\right)}
=\exp\left[2\sum_{m=1}\frac{N-N^{2m+1}}{2m+1}\zeta (2m+1)
\left(\frac{\rho}{2\pi i}\right)^{2m+1}
\right]\non
&&\qquad =1+2\zeta (3)\frac{c_3}{N}\left(\frac{\rho}{2\pi i}\right)^3+
{\cal O}(\rho^5)\,.\nom
\ea
The leading term in the $\hat{K}$ of the 
variety W is a 3rd Chern class of the M.
Generally the coefficients of $\hat{K}$ contain topological information
of M.
In fact, Chern classes of $c_{\ell}$ $(\ell =1,2,\cdots , N-2)$ 
of the manifold $M$ are generated by a 
function $c(\rho)$
\ba
&&c(\rho)=\frac{(1+\rho)^N}{1+N\rho}=1
+\sum_{\ell \geq  1}\rho^{\ell}\frac{c_{\ell}}{N}\,.\nom
\ea
Typical coefficients $X_{\ell}=N-N^{\ell}$ 
in the $\hat{K}$ are some combinations of 
Chern classes $c_{\ell}$
\ba
&& c(\rho)= 1+\sum_{\ell \geq  1}\rho^{\ell}\frac{c_{\ell}}{N}
= \exp\left(\sum_{\ell \geq 1}(-1)^{\ell -1}\rho^{\ell}\cdot 
\frac{X_{\ell}}{\ell}\right)\,.\nom
\ea
For examples, we list several $X_{\ell}$ for the Calabi-Yau case
\ba
&&X_1=0\,,\,\,X_2=-\frac{2}{N}c_2\,,\,\,X_3=\frac{3}{N}c_3\,,\,\,\non
&&X_4=-\frac{4}{N}c_4+\frac{2}{N^2}c_2^2
\,,\,\,X_5=\frac{5}{N}c_5-\frac{5}{N^2}c_3c_2\,.\,\,\nom
\ea
The series expansion Eq.(\ref{series}) converges around $z\sim 0$, that is, 
large complex structure point of the $W$.
For the purpose of an analytic continuation into the large complex
structure region, 
we find that the two sets of the periods are related by a transformation 
matrix $\tilde{M}$ with components $\tilde{M}_{k\ell}$
\ba
&&\tpi_k=\sum^{N-2}_{\ell =0}\tilde{M}_{k\ell}\Omega_{\ell}\,\,\,\,\,
(k=1,2,\cdots ,N-1)\,,\non
&&\tilde{M}_{k\ell}=(-N)\cdot (2\pi i)^{N-1}
\times \left[
 \sqrt{\hat{A}(\rho)}\cdot 
\frac{\alpha^k}{e^{\rho}-\alpha^k}\cdot (-\rho)^{\ell}
\right]\Biggr|_{\rho^{N-2}}\non
&&\qquad =
(-N)\cdot 
(2\pi i)^{N-1}\cdot \sum^{N-2}_{m =0}
G_{k,m}V_{m,\ell}\,,\non
&&G_{k,m}=\frac{-\alpha^k}{(\alpha^k -1)^{m+1}}\,\,\,\,\,
(1\leq k\leq N-1\,,\,\,0\leq m \leq N-2)\,,\non
&&\alpha =e^{\frac{2\pi i}{N}}\,,\non
&&V_{m,\ell}=\left[
\sqrt{\hat{A}(\rho)}\cdot (e^{\rho}-1)^m \cdot (-\rho)^{\ell}
\right]\Biggr|_{\rho^{N-2}}
\,\,\,\,\,(0\leq m\leq N-2\,,\,\, 0\leq \ell \leq N-2)\,.\nom
\ea
Here the transformation matrix $V$ contains
a square root of a topological invariant ``A-roof'' of the Calabi-Yau space
$M$
\ba
&&\hat{A}(\rho)=
\left(\frac{\dis \frac{\rho}{2}}{\dis \sinh \frac{\rho}{2}}\right)^N
\cdot\left(\frac{\dis \sinh \frac{N\rho}{2}}{\dis \frac{N\rho}{2}}\right)
=\frac{1}{\dis a(-\frac{\rho}{2\pi i}) a(+\frac{\rho}{2\pi i})}\non
&&\qquad 
=\exp\left[+\sum_{m=1}\frac{(-1)^mB_m}{(2m)!}\frac{N-N^{2m}}{2m}\rho^{2m}
\right]\non
&&\qquad =1+\frac{1}{12}\frac{c_2}{N}\rho^2 +{\cal O}(\rho^4)\,.\nom
\ea
The $B_m$s are Bernoulli numbers and are defined in our convention as
\ba
&&\frac{x}{e^x-1}=1-\frac{x}{2}-\sum_{n=1}^{\infty}
\frac{(-1)^n\cdot B_n}{(2n)!}x^{2n}\,,\non
&&B_1=\frac{1}{6}\,,\,\,B_2=\frac{1}{30}\,,\,\,
B_3=\frac{1}{42}\,,\,\,B_4=\frac{1}{30}\,,\cdots \,.\nom
\ea
Now we return to the {\kae} potential $K$.
By performing the analytic continuation 
of the Eq.(\ref{KO}), we can obtain a formula of the $K$
applicable in the large $\psi$ region. 
\ba
&&e^{-K}=
(-1)^N\left(\frac{2\pi i}{N}\right)^{N-2}\cdot\frac{1}{N^2}
\sum^{N-2}_{\ell ,\ell' =0}\left(V^{\dagger}{\cal I}V\right)_{\ell ,\ell'}
(-1)^{\ell +\ell'}\cdot \bar{\Omega}_{\ell}\Omega_{\ell'}\,,\nom
\ea
Here the ${\cal I}$ is a triangular matrix 
that combines $\bar{\Omega}$ and
$\Omega$ quadratically
\ba
&&{\cal I}_{m,m'}=
2^N\cdot i^N\cdot \sum^{N-1}_{k=1}
\frac{(-1)^k \left(\sin \frac{\pi k}{N}\right)^N}
{(\alpha^{-k}-1)^{m+1}(\alpha^k -1)^{m'+1}}=(-1)^m\delta_{m+m',N-2}
+\cdots \,.\nom
\ea
The ${\cal I}_{m,m'}$ has non-vanishing components only at
$m+m'\geq N-2$ ($m,m'=0,1,\cdots ,N-2$)
and we find an expression for the matrix $V^\dagger{\cal I}V$ 
\ba
&&\left(V^{\dagger}{\cal I}V\right)_{\ell ,\ell'}
=(-1)^{\ell'}\delta_{\ell +\ell' ,N-2}\,.\label{find}
\ea
We check validity of this equation Eq.(\ref{find}) concretely
for $N\leq 27$ and propose this formula for arbitrary $N(\geq 3)$ cases 
as a conjecture.

Finally we obtain a formula of the $K$ of $W$ with this equation
\ba
&&e^{-K}=(-1)^d\left(\frac{2\pi i}{N}\right)^{d}\cdot \frac{1}{N^2}
\left(\Omega^{\dagger}{\Sigma}\Omega\right)\,,\non
&&{\Sigma}_{\ell ,{\ell}'}
=(-1)^{\ell}\delta_{\ell +{\ell}' ,N-2}\,.\nom
\ea
The $e^{-K}$ is constructed by combining a holomorphic $d$ form and
an anti-holomorphic one quadratically. Both parts are decomposed 
by a dual basis of (real) homology cycles and  
their coefficients are realized as periods.  
Then we can understand that a matrix which
combines the periods and their complex conjugates is
an intersection matrix of the cycles.
In our case, 
the ${\Sigma}$ is an intersection matrix of 
homology cycles associated with the set of periods $\Omega_m$ of $W$. 
The result means that the cycles we used here are 
combined into a symplectic $\mbox{USp}(d+1)$ or 
an $\mbox{SO}(\frac{d}{2}+1,\frac{d}{2})$ invariant bases for
respectively $d=$odd or  $d=$even cases.
But the basis is not an integral one and we have to perform
an appropriate  linear transformation with 
fractional rational numbers to construct
a canonical basis of a central charge.
More details will appear in our next paper.

Let us study behaviors of a metric, a curvature in the large $\psi$
region of $W$. 
Powers of logarithm of $\psi$ appear in 
the {\kae} potential $K$ 
in the leading expansion
\ba
&&e^{-K}=\frac{1}{d!N^2}\{2\log(N|\psi|)\}^d 
\times \left[1+\sum^{d}_{n=1}\frac{d!}{(d-n)!}\cdot 
\frac{(-1)^n\cdot \hat{K}_n}{\{2N\log (N|\psi|)\}^n}\right]+\cdots \non
&&\qquad =\frac{1}{d!N^2}\{2\log(N|\psi|)\}^d 
\times \Bigl[
1+4\left(\matrix{d\cr 3}\right)(N^3-N)\zeta (3)\cdot
\{2N\log (N|\psi|)\}^{-3}\non
&&\qquad +48
\left(\matrix{d\cr 5}\right)(N^5-N)\zeta (5)\cdot
\{2N\log (N|\psi|)\}^{-5}+\cdots 
\Bigr]\,.\nom
\ea
Here the $\hat{K}_n$s are coefficients of the series expansion of the
$\hat{K}$
\ba
&&\hat{K}(\rho)=1+\sum_{n=3}\hat{K}_n
\left(\frac{\rho}{2\pi i}\right)^n\,,\non
&&\hat{K}_3=-\frac{2}{3}(N^3-N)\zeta (3) =\frac{2}{N}c_3\zeta
(3)\,,\non
&&\hat{K}_5=-\frac{2}{5}(N^5-N)\zeta (5) =\frac{2}{N}
(c_5-\frac{1}{N}c_3\cdot c_2)\zeta
(5)\,,\nom
\ea 
They are related to the Chern classes of the $d$-fold $M$.
Also there appear 
$\zeta (2m+1)$s ($m=1,2,\cdots $ with $2m+1\leq d$)
, which might be transcendental numbers, 
in this formula. We summarize several concrete examples of the 
$\hat{K}_n$ in the appendix.
The exponent of power of the logarithm is at most $d$ for the $d$-fold.
In addition, there are parts of infinite series with respect to the
 $\psi$ in the $e^{-K}$. They are omitted as an 
abbreviated symbol ``$\cdots$''.

Next we calculate the {\kae} metric of this B-model associated with $W$
\ba
&&g_{\psi\bar{\psi}}=
\frac{d}{\{2|\psi|\log (N|\psi|)\}^2}\times 
\Bigl[
1-16\left(\matrix{d-1\cr 2}\right)(N^3-N)\zeta (3)\cdot
\{2N\log (N|\psi|)\}^{-3}\non
&&\qquad -288\left(\matrix{d-1\cr 4}\right)(N^5-N)\zeta (5)\cdot
\{2N\log (N|\psi|)\}^{-5}+\cdots 
\Bigr]\,.\nom
\ea
Also we obtain the
scalar curvature in this large $\psi$ region
\ba
&&R=-\frac{4}{d}\times 
\Bigl[1-80\left(\matrix{d-1\cr 2}\right)(N^3-N)\zeta (3)\cdot
\{2N\log (N|\psi|)\}^{-3}\non
&&\qquad 
-4032\left(\matrix{d-1\cr 4}\right)(N^5-N)\zeta (5)\cdot
\{2N\log (N|\psi|)\}^{-5}+\cdots 
\Bigr]\,.\nom
\ea
In the $|\psi|=\infty$ limit, the curvature is a negative constant and 
its absolute value is inversely proportional to the dimension $d$.
The $|R|$ decreases
 apart from the point $|\psi|=\infty$ 
for $N\geq 5$ cases.
The leading term of the corrections in the brackets is
inversely proportional to the $\{\log (N\psi)\}^{-3}$ and contains 
$\zeta (3)$ as its coefficient.
For the $N=3,4$ cases, the associated curvatures $R$s are constants except for 
points in the $\psi$-plane with $\psi^N=1$.

Next we discuss an invariant coupling.
The {\kae} potential is not a function but a section of a line bundle
and
there is an arbitrariness of multiplications by  
holomorphic and anti-holomorphic functions.
Also a $d$-point correlation function in the B-model is a section with a 
weight $2$
\ba
\kei_{\underbrace{\scriptstyle\psi\psi\cdots \psi}_{d\,
\mbox{\scriptsize times}}}
=\frac{1}{N^d}\cdot \frac{N\psi^2}{1-\psi^N}\,.\nom
\ea
But there is an invariant $d$-point function $\kei$ that is constructed 
by combining the metric, the $K$ and $\kei_{\psi \cdots \psi}$
\ba
\kei =(g_{\psi\bar{\psi}})^{-d/2}e^{K}
|\kei_{\psi\psi\cdots \psi}|\,.\nom
\ea
In our normalization, it is expressed in the large $|\psi|$ limit as
\ba
&&\kei=
\frac{d!}{d^{d/2}\cdot N^{d-3}}\times\Bigl[
1+20 \left(\matrix{d\cr 3}\right)(N^3-N)\zeta (3)\cdot 
\{2N\log (N|\psi|)\}^{-3}\non
&&\qquad +672\left(\matrix{d\cr 5}\right)(N^5-N)\zeta (5)\cdot 
\{2N\log (N|\psi|)\}^{-5}+\cdots
\Bigr]\,.\nom
\ea
Up to coefficients, the corrections have the same structures for the 
metric, the $R$ and the $\kei$.
In the small complex structure limit (at $\psi =0$), 
this invariant coupling is 
evaluated as
\ba
\kei =\left[\frac{\dis \Gam\!\left(\frac{1}{N}\right)}
{\dis \Gam\!\left(1-\frac{1}{N}\right)}\right]^{\frac{1}{2}N(d-2)}
\left[\frac{\dis \Gam\!\left(1-\frac{2}{N}\right)}
{\dis \Gam\!\left(\frac{2}{N}\right)}\right]^{\frac{1}{2}Nd}
\times \frac{1}{N^{d-3}}\,.\nom
\ea
For an example, we write down the result in the  $d=3$ case 
\ba
\kei=\frac{2}{\sqrt{3}}\left[
1-12c_3\zeta (3)\cdot \{10\log (5|\psi|)\}^{-3}+\cdots
\right]\,,\,\,\,c_3=-200\,.\nom
\ea
This shows that the coupling increases apart from the $|\psi|=\infty$.

Also we compare the leading value of this $\kei$ in the small $|\psi|$
with that in the large  $|\psi|$ limits
\ba
&&\mbox{small $\psi$}\,;\,\kei =
\left(\frac{\Gam\!(\frac{1}{5})}{\Gam\!(\frac{4}{5})}\right)^{5/2}
\left(\frac{\Gam\!(\frac{3}{5})}{\Gam\!(\frac{2}{5})}\right)^{15/2}\non
&&\qquad =1.55531898996323897249994495854237822\cdots \non
&&\mbox{large $\psi$}\,;\,\kei =\frac{2}{\sqrt{3}}\non
&&\qquad =1.154700538379251529018297561003914911\cdots \,.\nom
\ea
The $\kei$ in the small $\psi$ case can be analyzed by using our previous 
result in \cite{KS7}. 
The coupling in the small $\psi$ region is stronger than that in the
large $\psi$ region in an amount of 34.7\%.

\section{Large Radius Region}

In the previous section, we study properties of physical quantities in 
the large complex structure point of the $d$-fold $W$. 
It is known that the large complex structure point is related to the 
large radius point of the partner $M$.
It is possible to translate the results of $W$ to those of the $M$. 
For the purpose of this program, we introduce a mirror map $t$
\ba
&&2\pi i t
=\log z+\frac{\dis\sum^{\infty}_{n=1}\frac{(Nn)!}{(n!)^N}
\left(\sum^{Nn}_{m=n+1}\frac{N}{m}\right)z^n}
{\dis \sum^{\infty}_{n=0}\frac{(Nn)!}{(n!)^N}z^n}\,,\,\,\,
q=e^{2\pi i t}\,,\non
&&x_n=\frac{1}{(2\pi i)^n}\frac{1}{n!}\del^n_{\rho }\log
\left[\sum^{\infty}_{m=0}\frac{\Gam\!(N(m+\rho)+1)}
{\Gam\!(N\rho +1)}\left(\frac{\Gam\!(\rho +1)}
{\Gam\!(m+\rho +1)}\right)^N z^m\right]\Biggr|_{\rho =0}\non
&&\qquad =
\sum_{m=1}^{\infty} {a_{n,m}}{q^m}\,\,\,\,(n\geq 2)\,,\non
&&a_{n,1}=N!\cdot S_n(\beta_1(1),\cdots ,\beta_n(1))\,,\non
&&a_{n,2}=\frac{(2N)!}{2^N}\cdot S_n(\beta_1(2),\cdots ,\beta_n(2))\non
&&\qquad -(N!)^2 \left(\sum^N_{m=2}\frac{N}{m}\right)\cdot 
S_n(\beta_1(1),\cdots ,\beta_n(1))\non
&&\qquad 
-\frac{1}{2}(N!)^2\cdot S_n(2\beta_1(1),\cdots ,2\beta_n(1))\,,\non
&&\beta_m(n)=N^m\cdot \sum^{Nn}_{k=1}\frac{(-1)^{k-1}(k-1)!}{k^m}-
N\cdot \sum^{n}_{k=1}\frac{(-1)^{k-1}(k-1)!}{k^m}\,.\nom
\ea
This $t$ is a coordinate of the {\kae} moduli space or the 
coefficient of the complexified {\kae} form
\ba
B+iJ=t[D]\,,\,\,\,\,[D]\in \mbox{H}^2(M)\,.\nom
\ea
The $[D]$ is a Poincar\'e dual of a divisor ``$D$'' of the $M$.
The $S_n(x_1,x_2,\cdots ,x_n)$s are Schur polynomials and are defined as
\ba
\exp\left(\sum_{m=1}x_m u^m\right)=\sum_{n=0}S_n(x_1,x_2,\cdots
,x_n)u^n\,.\nom
\ea
Next we rewrite the $K$ in this coordinate 
\ba
&&e^{-K}=(-1)^d\left(\frac{2\pi i}{N}\right)^d \cdot \frac{1}{N^2}
(\Omega^\dagger \Sigma \Omega)\,,\non
&&\left(\Omega^{\dagger}{\Sigma}\Omega\right)
=\sum^{N-2}_{\ell =0}\bar{\Omega}_{\ell}\cdot (-1)^{\ell}
\Omega_{N-2-\ell}=
\left[
\hat{K}(\rho)\cdot 
\overline{\vpi\left(-\frac{\rho}{2\pi i}\,;\,z\right)}\cdot 
\vpi\left(+\frac{\rho}{2\pi i}\,;\,z\right)
\right]\Biggr|_{\rho^{N-2}}\non
&&\qquad 
=|\vpi_0|^2\times 
\left[
\hat{K}(\rho)e^{\rho (t-\bar{t})}
\exp\left(\sum_{n\geq 2}\rho^n (x_n+(-1)^n\bar{x}_n)
\right)
\right]\Biggr|_{\rho^{N-2}}\,.\label{find2}
\ea
Here the first term in the brackets contains 
characteristic classes of the $M$.
Its coefficients in the series expansion 
are represented as some combinations of Chern classes of the $M$.
Also there appear Riemann's zeta functions evaluated at positive odd
integers. These are irrational numbers and might be transcendental numbers.
It implies some arithmetic properties of this model.
Next the second term is associated to the imaginary part of the $t$.
When we translate the formal parameter $\rho$ into a divisor ``$D$'' of
the hyperplane, the $\rho (t-\bar{t})$ is identified with the 
{\kae} form $J$ of $M$
\ba
\rho (t-\bar{t})=2i\rho\im (t)\leftrightarrow 
2i\,\mbox{Im}(t)[D]=2iJ\,.\nom 
\ea
This second term contains only imaginary part of $t$ and is invariant
under an arbitrary shift of 
$\mbox{Re}(t)\rightarrow \re (t)+{\bf a}$ for ${}^\forall {\bf a}\in
{\bf R}$. This symmetry is a classical one and is broken at the quantum level. 
In fact, the third term in Eq.(\ref{find2}) in not invariant under 
this arbitrary shift because the $x_n$s are expressed 
as series expansions of the variable $q=e^{2\pi i t}$.
This 3rd term is invariant under only
integral shifts of $t\rightarrow  t+n$
with $n\in {\bf Z}$.
This term contains information about 
 non-perturbative effects of the worldsheet instantons.
They break the Peccei-Quinn symmetry into a symmetry under the 
integral shift of the $\mbox{Re}(t)$. 
The real part of the $t$ is related to the 2nd rank antisymmetric
field $B$ in the NS-NS sector
\ba
\mbox{Re}(t)[D]=B\,,\nom
\ea
and the shift of the $\re (t)\rightarrow \re (t)+n$ is equivalent to a
shift of the $B$ field, $B\rightarrow B+[D]$.

\begin{figure}[htbp]
\epsfxsize=8cm
 \centerline{\epsfbox{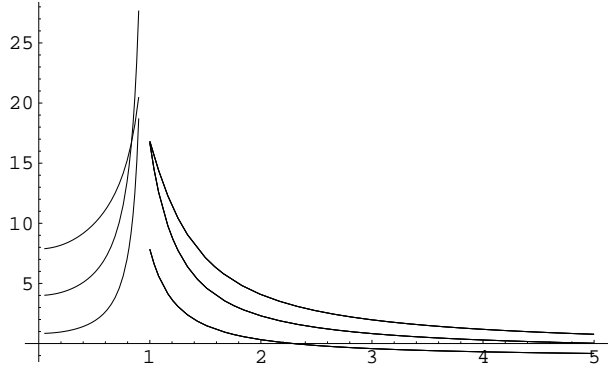}}
\epsfxsize=8cm
  \caption{Scalar Curvature $R$ for $N=5,6,7$ cases. The axis of
 abscissa represents the value of the $\psi\in {\bf R}$. The 
axis of ordinate  corresponds to the scalar curvature $R$.
The point $\psi =1$ is a singular point where associated scalar
 curvatures blow up. 
The $R$ vanishes at some point $\psi =\psi_0$
 for each $N$.
In the range $\psi >\psi_0$, the $R$ is negative and its 
asymptotic value is $-4/d$.}
\label{r} 
\end{figure}

Finally we will write down our result for the $K$
\ba
&&e^{-K}=
(-1)^d\left(\frac{2\pi i}{N}\right)^d\cdot \frac{1}{N^2}|\vpi_0|^2
\cdot S_d(\tilde{y}_1,\tilde{y}_2,\cdots ,\tilde{y}_d)\non
&&\qquad =(-1)^d\left(\frac{2\pi i}{N}\right)^d\cdot \frac{1}{N^2}|\vpi_0|^2
\cdot \frac{(t-\bar{t})^d}{d!}\times 
\left[1+\sum^{d}_{\ell =2}\left(\matrix{d\cr \ell}\right)
\frac{\ell ! S_{\ell}(0,\tilde{y}_2,\cdots ,\tilde{y}_{\ell})}
{(t-\bar{t})^{\ell}}
\right]\,,\non
&&\hat{k}_{2n+1}:=2\cdot \frac{N-N^{2n+1}}{2n+1}\cdot
\frac{\zeta (2n+1)}{(2\pi i)^{2n+1}}\,\,\,\,\,\,\,(n\geq 1)\,,\non
&&\hat{K}(\rho)=\exp\left(\sum_{m\geq 1}\hat{k}_{2m+1}\cdot\rho^{2m+1}\right)
=1+\sum_{n\geq 1}\hat{K}_n\cdot \left(\frac{\rho}{2\pi i}\right)^n\,,\non
&&y_n:=x_n+(-1)^n\bar{x}_n\,\,\,\,(n\geq 2)\,,\non
&&\tilde{y}_n=
\left\{
\matrix{\dis t-\bar{t} & (n=1)\cr
y_{2m} & (n=2m\,;\,m\geq 1)\cr
y_{2m+1}+\hat{k}_{2m+1} & (n=2m+1\,;\,m\geq 1) }
\right.\,.\nom
\ea
The Schur functions contain 
``loop''\footnote{The $\frac{c_3\zeta (3)}{(2\pi i)^3}$ is interpreted
as an effect of  
loop corrections at the 4-loop perturbative calculation of the 
sigma model with 3 dimensional Calabi-Yau target spaces. We do not
know that the other 
terms $\hat{K}_{n}$s 
in the $\hat{K}$ can be interpreted directly as perturbative loop 
corrections of the sigma model at higher loop calculations.} 
and non-perturbative effects of the 
model
\ba
&&S_2=2\re (x_2)\,,\,\,S_3=2i\im (x_3)+\hat{k}_3\,,\non
&&S_4=2[\re (x_4)+(\re (x_2))^2]\,,\non
&&S_5=2i\im (y_5)+4i\re (x_2)\im (x_3)+\hat{k}_5+
2\re (x_2)\cdot \hat{k}_3\,.\nom
\ea
Especially we obtain an asymptotic formula of the 
$y_n$s in the range of the large radius 
volume 
\ba
&&\left\{
\matrix{\tilde{y}_{2m}\rightarrow 0 & (m\geq 1)\cr
\tilde{y}_{2m+1}\rightarrow \hat{k}_{2m+1} & (m\geq 1)}
\right.\non
&&S_{\ell}(0,\tilde{y}_2,\cdots ,\tilde{y}_{\ell})
\rightarrow \hat{K}_{\ell}\,\,\,\,(\ell \geq 2)\,.\nom
\ea
In this limit, we can neglect non-perturbative corrections and write down 
the $K$, a metric and a scalar curvature of the moduli space of $M$ as power
series of the $(t-\bar{t})$
\ba
&&e^{-K}=(-1)^d\left(\frac{2\pi i}{N}\right)^d\frac{1}{N^2}
\frac{1}{d!}(t-\bar{t})^d
\cdot \left[1+\sum^{d}_{\ell =3}\left(\matrix{d\cr \ell}\right)
\frac{\ell !\hat{K}_{\ell}}{(2\pi i)^{\ell}\cdot (t-\bar{t})^{\ell}}\right]
\,,\non
&&g_{t\bar{t}}=\frac{-d}{(t-\bar{t})^2}
\cdot \Biggl[
1 - {\frac{72\,\left(\matrix{d\cr
        3}\right)\,\hat{K}_3\,}{(2\pi i)^3\cdot 
d\cdot\,{{\left( t - \bar{t} \right)
        }^3}}}-
{\frac{3600\,\left(\matrix{d\cr 5}\right)\,\hat{K}_5\,}
{(2\pi i)^5\cdot d\cdot\,{{\left( t - \bar{t} \right) }^5}}} + \cdots\Biggr]
\,,\non
&&R_{t\bar{t}}=-\del_t\del_{\bar{t}}\log g_{t\bar{t}}=
\frac{2}{(t-\bar{t})^2}
\cdot \Biggl[
1 - {\frac{432\,\left(\matrix{d\cr
        3}\right)\,\hat{K}_3\,}
{(2\pi i)^3\cdot d\cdot\,{{\left( t - \bar{t} \right)
        }^3}}}-
{\frac{54000\,\left(\matrix{d\cr 5}\right)\,\hat{K}_5\,}
{(2\pi i)^5\cdot d\cdot\,{{\left( t - \bar{t} \right) }^5}}} + \cdots
\Biggr]
\,,\non
&&R=2g^{t\bar{t}}R_{t\bar{t}}=-\frac{4}{d}
\cdot\Biggl[
1 - {\frac{360\,\left(\matrix{d\cr 3}\right)
\,\hat{K}_3\,}{(2\pi i)^3\cdot d\cdot\,{{\left( t - \bar{t} \right) }^3}}}
-{\frac{50400\,\left(\matrix{d\cr 5}\right)\,\hat{K}_5\,}
{(2\pi i)^5\cdot d\cdot\,{{\left( t - \bar{t} \right) }^5}}} + \cdots
\Biggr]\,.\nom
\ea
The line element of the A-model moduli space 
is given as $ds^2=g_{t\bar{t}}dtd\bar{t}$.
The large $\im (t)$ limit, the metric and the Ricci tensor are 
inversely proportional to the $(t-\bar{t})$.
But apart from the point $\im (t)=\infty$, 
there appear terms $\im (t)^n$ ($n >2$) and also non-perturbative 
corrections with $e^{2\pi i\,n\,t }$ or $e^{-2\pi i\,n\,\bar{t} }$
($n\geq 1$) for $d\geq 3$ cases.
The large $\im (t)$ point corresponds to the large $\psi$ point. 
On the other hand, a related value of the ``$t$'' 
at the $\psi =0$ point does not
vanish, but it is finite
\ba
t(\psi =0)=-\frac{1}{2}+\frac{i}{2\tan \frac{\pi}{N}}\,.\nom
\ea
At the $\psi =0$ point, the $R$, $R_{t\bar{t}}$ and $g_{t\bar{t}}$ do 
not vanish and are evaluated as
\ba
&&g_{t\bar{t}}=\left(2\sin \frac{2\pi }{N}\right)^2
\left(2\cos \frac{\pi}{N}\right)^{-N+2}\,,\non
&&R=-4+2\left[\frac{\Gam\!\left(\frac{1}{N}\right)
\Gam\!\left(\frac{3}{N}\right)}
{\Gam\!\left(1-\frac{1}{N}\right)
\Gam\!\left(1-\frac{3}{N}\right)}\right]^N\cdot 
\left[\frac{\Gam\!\left(1-\frac{2}{N}\right)}
{\Gam\!\left(\frac{2}{N}\right)}\right]^{2N}\,,\non
&&R_{t\bar{t}}=\frac{1}{2}g_{t\bar{t}}R\,.\nom
\ea
The scalar curvature is 
positive around the Gepner point for $d\geq 3$.
When we increase the $\psi$ from zero to one,
the $R$ increases monotonically with the $\psi$ for each $N$.
The point $\psi =1$ is a singular point where associated scalar
 curvatures blow up. When one passes through the $\psi =1$ and 
increases the value of the $\psi$, the associated 
$R$ decreases monotonically and vanishes at some point $\psi =\psi_0$
 for each $N$.
In the range $\psi >\psi_0$, the $R$ is negative and its 
asymptotic value is $-4/d$.
That is to say, it is 
negative in the large $\psi$ region for 
$d\geq 3$.
The behaviors of the curvatures for $N=5,6,7$ are shown in Fig.\ref{r} 

\begin{figure}[htbp]
\epsfxsize=8cm
 \centerline{\epsfbox{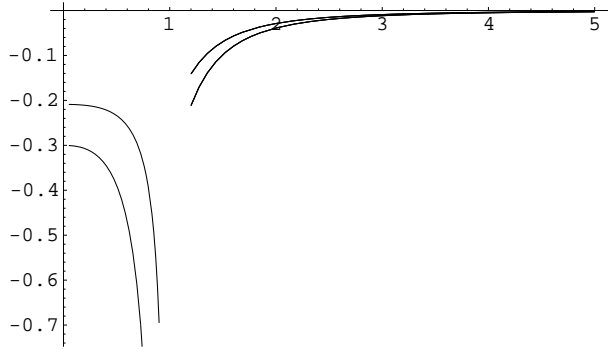}}
\epsfxsize=8cm
  \caption{Ricci Tensor $R_{\psi\bar{\psi}}$ for torus ($N=3$) and K3 ($N=4$) 
 cases. The axis of abscissa represents 
the value of the $\psi\in {\bf R}$. The 
axis of ordinate  corresponds to the Ricci Tensor $R_{\psi\bar{\psi}}$.
The point $\psi =1$ is a singular point where the $R_{\psi\bar{\psi}}$s
blow up. But the scalar curvatures are constants except for a point
 $\psi =1$.}
\label{ric} 
\end{figure}

For the torus case, the curvature is 
$R=-4$ both in the small $\psi$ and in the large $\psi$ regions. 
Similarly, for the K3 case, the $R$ is $(-2)$ in the two limits. 
Associated Ricci tensors at the $\psi =0$ are obtained for $N=3,4$ 
respectively
\ba
&&R_{\psi\bar{\psi}}=-18\cdot 
\left[\frac{\Gam\!\left(\frac{2}{3}\right)}
{\Gam\!\left(\frac{1}{3}\right)}\right]^6
=-0.30021677774546778674\cdots \,,\,\,\,(\mbox{$N=3$ case})\,,\non
&&R_{\psi\bar{\psi}}=-16\cdot 
\left[\frac{\Gam\!\left(\frac{3}{4}\right)}
{\Gam\!\left(\frac{1}{4}\right)}\right]^4
=-0.20880017792822456419\cdots\,,\,\,\,(\mbox{$N=4$ case})\,.\nom
\ea
When we increase the $\psi$ from zero to one,
the $R_{\psi\bar{\psi}}$ 
decreases monotonically with the $\psi$ for each $N$.
The point $\psi =1$ is a singular point where the $R_{\psi\bar{\psi}}$s
blow up.  When one passes through the $\psi =1$ and 
increases the value of the $\psi$, the associated 
$R_{\psi\bar{\psi}}$ decreases monotonically
 for each $N$ as shown in Fig.\ref{ric}. 
But the scalar curvatures are constants except for a point
 $\psi =1$.

\section{Two-Point Functions}
The other constituent blocks of the topological model are three-point
couplings $\{\kei_{\ell}\}$ and two-point functions. Each A-model
operator ${\cal O}^{(\ell)}$ is associated with a cohomology element 
$e_{\ell}\in \mbox{H}^{2\ell}$(M). The $\kei_{\ell}$ is a fusion
coupling of ${\cal O}^{(1)}$ and ${\cal O}^{(\ell)}$
\ba
&&{\cal O}^{(1)}\cdot {\cal O}^{(\ell)}=
  \mbox{\Large \mbox{$\kappa$}}_{\ell} 
{\cal O}^{(\ell +1)}\,\,\,\,(\ell =0,1,\cdots ,d-1)\,.\non
&&{\cal O}^{(1)}\cdot {\cal O}^{(d)}=0\,,\non
&&\mbox{\Large\mbox{$\kappa$}}_{0}=1\,,\,\,\,\non
&&\mbox{\Large\mbox{$\kappa$}}_{\ell}=
\del \frac{1}{ \kei_{\ell -1}}\del \frac{1}{ \kei_{\ell -2}}
\del \cdots \del \frac{1}{ \kei_{1}}\del \frac{1}{ \kei_{0}}\del 
S_{\ell +1}(t,x_2,x_3,\cdots ,x_{\ell +1})\non
&&\qquad =1+{\cal O}(q)\,,\,\,\,\,\,
(1\leq \ell \leq d-1)\,.\nom
\ea
Also the topological metric $\mbox{\large\mbox{$\eta$}}_{\ell ,m}$ is 
given as
\ba
\mbox{\large\mbox{$\eta$}}_{\ell ,m}=\langle {\cal O}^{(\ell)}
{\cal O}^{(m)}\rangle =N\delta_{\ell +m,d}\,.\nom
\ea
On the other hand, the hermitian two-point functions 
$\langle\bar{\cal O}^{(\ell)}|{\cal O}^{(m)}\rangle$ are calculated by using 
a method of the $tt^{\ast}$-fusion. In our case, the correlators have
diagonal forms and are given as \cite{KS5}
\ba
&&\langle\bar{\cal O}^{(\ell)}|{\cal O}^{(m)}\rangle =
e^{q_{\ell}}\delta_{\ell ,m}\,\,\,\,\,\,(0\leq \ell \leq d\,,\,
0\leq m\leq m)\,,\non
&&\tilde{q}_0=q_0\,,\non
&&\tilde{q}_{\ell}=q_{\ell}+\sum_{n=0}^{\ell -1}\log 
|\mbox{\Large{\mbox{$\kappa$}}}_n|^2\,\,\,\,(\ell \geq 1)\,,\non
&&\del\bar{\del}\tilde{q}_0+e^{\tilde{q}_1-\tilde{q}_0}=0\,,\non
&&\del\bar{\del}\tilde{q}_{\ell}+e^{\tilde{q}_{\ell +1}-\tilde{q}_{\ell}}
-e^{\tilde{q}_{\ell}-\tilde{q}_{\ell -1}}
=0\,\,\,(1\leq \ell\leq d-1)\,,\non
&&\del\bar{\del}\tilde{q}_d-e^{\tilde{q}_d-\tilde{q}_{d-1}}=0\,.\nom
\ea
In the large radius limit, the $\tilde{q}_{\ell}$s behave as
\ba
&&e^{\tilde{q}_{0}}=e^{-K}=(-1)^d\left(\frac{2\pi i}{N}\right)^d\cdot
\frac{1}{N^2}\cdot \frac{(t-\bar{t})^d}{d!}+\cdots\,,\non
&&e^{\tilde{q}_{\ell +1}-\tilde{q}_{\ell }}=
-\frac{(\ell +1)(d-\ell)}{(t-\bar{t})^2}+\cdots \,\,\,\,\,(0\leq \ell
\leq d-1)\,.\nom
\ea
In this case, any corrections are suppressed and 
normalized two-point functions are inversely proportional
to the $\im (t)^{2\ell}$ ($\ell =0,1,2,\cdots ,d$)
\ba
\frac{\dis \langle \bar{{\cal O}}^{(\ell)}|{\cal O}^{(\ell)}\rangle}
{\dis \langle \bar{{\cal O}}^{(0)}|{\cal O}^{(0)}\rangle}=
e^{\tilde{q}_{\ell }-\tilde{q}_{0}}=
(-1)^{\ell}(\ell !)^2\left(\matrix{d\cr \ell}\right)
\frac{1}{(t-\bar{t})^{2\ell}}+\cdots \,\,\,\,(0\leq \ell\leq d)\,.\nom
\ea
In the finite $\im (t)$ case, there appear corrections in power series and 
non-perturbative corrections of exponential types. Those are
abbreviated as a symbol ``$\cdots$'' in the above formula.
Even in that generic case, these $q_{\ell}$s 
are described by combining the curvature $R$, the 
metric $g_{t\bar{t}}$ and the $K$
\ba
&&e^{\tilde{q}_0}=e^{-K}\,,\,\,\,
e^{\tilde{q}_1-\tilde{q}_0}=g_{t\bar{t}}\,,\,\,\,
e^{\tilde{q}_2-\tilde{q}_1}=g_{t\bar{t}}
\left(\frac{R}{2}+2\right)\,,\non
&&e^{\tilde{q}_3-\tilde{q}_2}=
g_{t\bar{t}}
\left[3\left(\frac{R}{2}+1\right)
-g^{t\bar{t}}\del_{t}\bar{\del}_{\bar{t}}
\log\left(\frac{R}{2}+2\right)\right]\,,\non
&&e^{\tilde{q}_4-\tilde{q}_3}=
g_{t\bar{t}}
\Biggl[4+3R-2 g^{t\bar{t}}\del_{t}\bar{\del}_{\bar{t}}
\log\left(\frac{R}{2}+2\right)\non
&&\qquad \qquad 
-g^{t\bar{t}}\del_{t}\bar{\del}_{\bar{t}}
\log\Bigl[
3\left(\frac{R}{2}+1\right)
-g^{t\bar{t}}
\del_{t}\bar{\del}_{\bar{t}}
\log\left(\frac{R}{2}+2\right)\Bigr]\Biggr]\,,\non
&&\qquad \cdots\,.\nom
\ea
We know the formula of the $K$, $R$, and $g_{t\bar{t}}$
and can evaluate moduli dependences of these correlators.

\section{Quintic}
In this section, we investigate the $K$ for 
a quintic $M$ case and compare our results with
those by Candelas et al\cite{CDGP}.
The set of periods $\{\Omega_{\ell}\}$
of an associated mirror $W$ is expressed by using the $\varpi$
\ba
&&\vpi\left(\frac{\rho}{2\pi i}\right)\sqrt{\hat{K}(\rho)}=:
\sum_{\ell \geq 0}\rho^{\ell}\Omega_{\ell}\,,\non
&&\sqrt{\hat{A}(\rho)}=1+\frac{5}{12}\rho^2\,,\,\,\,
\sqrt{\hat{K}(\rho)}=1-\frac{40}{(2\pi i)^3}\zeta (3)\rho^3\,,\non
&&\hat{c}=\frac{40}{(2\pi i)^3}\zeta (3)\,,\non
&&\Omega =\left(\matrix{\Omega_0\cr\Omega_1\cr\Omega_2\cr\Omega_3}\right)
=\left(\matrix{1\cr t\cr \frac{1}{2}t^2 +S_2(0,x_2)
\cr\frac{1}{6}t^3-\hat{c}+tS_2(0,x_2)+S_3(0,x_2,x_3)
}\right) \,.\,\label{ap1}
\ea
The prepotential $F$ of $M$ is expressed as a sum of 
a polynomial part of $t$ and a non-perturbative part $f$
\ba
&&F=-\frac{\kappa}{6}t^3+\frac{1}{2}at^2+bt+\frac{1}{2}c+f\,,\non
&&a=\frac{-11}{2}\,,\,\,b=\frac{25}{12}\,,\,\,c=\frac{c_3
\zeta (3)}{(2\pi i)^3}=-5\hat{c}
\,,\,\,c_3 =-200\,.\nom
\ea
The effects of the instantons are encoded in this function $f$. 
The $F$ leads to a   a canonical  set of 
basis $\{\Pi_{\ell}\}$. The $\Pi_{\ell}$s
are represented as some linear combinations of 
periods
\ba
&&\Pi =\left(\matrix{\Pi_0\cr\Pi_1\cr\Pi_2\cr\Pi_4}\right)
=\left(\matrix{1\cr t\cr \del_tF\cr t\del_tF-2F}\right)
={\cal N}\Omega\,,\,\,\,\label{ap2}\\
&&{\cal N}=\left(\matrix{1&0&0&0\cr 0&1&0&0\cr {b}&{a}&-\kappa&0\cr 
0 & -{b}&0 &-\kappa }\right)
=\left(
\matrix{ 1 & 0 & 0 & 0 \cr 0 & 1 & 0 & 0 \cr {\frac{25}{12}} & -{\frac{11}
     {2}} & -5 & 0 \cr 0 & -{\frac{25}{12}} & 0 & -5 \cr  } 
\right)
\,,\non
&&\kappa =5\,,\,\,a=-11/2\,,\,\,b=25/12\,.\,\,\,\nom
\ea
By comparing these two approaches Eqs.(\ref{ap1}),(\ref{ap2}), 
we obtain relations for the $c$ and the $f$
\ba
&&c=-\kappa\hat{c}=\frac{c_3\zeta (3)}{(2\pi i)^3}\,,\non
&&f=\frac{\kappa}{2}S_3(0,x_2,x_3)\,,\non
&&\del_t f=-\kappa S_2(0,x_2)\,.\nom
\ea
Then the {\kae} potential $K$ in the A-model 
is evaluated by using the $F$ up to 
a overall normalization factor with a suitable choice of a section of 
an associated line bundle
\ba
&&e^{-K}=(t-\bar{t})(\del F+\bar{\del}\bar{F})-2(F-\bar{F})\non
&&\qquad =-\frac{\kappa}{6}(t-\bar{t})^3+(\bar{a}-a)t\bar{t}
-(b-\bar{b})(t-\bar{t})-(c-\bar{c})
+(t-\bar{t})(\del f+\bar{\del}\bar{f})-2(f-\bar{f})\non
&&\qquad =-\frac{\kappa}{6}(t-\bar{t})^3-2c
+(t-\bar{t})(\del f+\bar{\del}\bar{f})-2(f-\bar{f})\,,\non
&&\qquad =-\kappa\left[\frac{1}{6}(t-\bar{t})^3-2\hat{c}
+(t-\bar{t})(S_2+\bar{S}_2)+(S_3-\bar{S}_3)
\right]\,\non
&&\qquad =(-1)^3\int_{\mbox{\scriptsize CY}_3}
\left[\hat{K}(\rho)\cdot \overline{\varpi(-\frac{\rho}{2\pi i})}
\cdot {\varpi(+\frac{\rho}{2\pi i})}
\right]\Biggr|_{\rho =[D]}\,,\non
&&\int_{\mbox{\scriptsize CY}_3} [D]\cdot [D]\cdot [D]=\kappa\,.\nom
\ea
Here we used the facts the $a$ and $b$ are real numbers and the $c$ 
is a pure imaginary number.
This formula coincides with our formula Eq.(\ref{find2}). Now we make a 
remark here: The basis $\Omega$ does 
not coincide with the $\Pi$, but it is related
to the $\Pi$ by a kind of a symplectic transformation, that is represented as a
matrix ${\cal N}$
\ba
&&\Pi={\cal N}\Omega\,,\,\,\,
{\cal N}^t\Sigma {\cal N}=(-5)\cdot \Sigma\,.\nom
\ea
It leads to the same $K$ up to a 
multiplicative factor because the following relation is satisfied
\ba
e^{-K}\sim \Pi^\dagger \Sigma\Pi =\Omega^\dagger {\cal N}^t\Sigma 
{\cal N}\Omega 
=(-5)\cdot \Omega^\dagger \Sigma \Omega\,.\nom
\ea

\section{Generalization}
In this section, we propose a formula of the $K$ by generalizing the 
previous result of the Fermat type.
We consider complete intersections $M$ 
of $\ell$ hypersurfaces $\{p_{j}=0\}$ in
products of $k$ projective spaces ${\cal M}$
\ba
M:=\left(\matrix{{\bf P}^{n_1}(w_{1}^{(1)},\cdots ,w_{n_1+1}^{(1)})\cr
\vdots \cr
{\bf P}^{n_k}(w_{1}^{(k)},\cdots ,w_{n_1+1}^{(k)})
}
\right|\!\!
\left|
\matrix{d_{1}^{(1)}\cdots d_{\ell}^{(1)}\cr
\vdots \cr
d_{1}^{(k)}\cdots d_{\ell}^{(k)}}
\right)\,.\nom
\ea
The $d^{(i)}_j$ are degrees of the coordinates of 
${\bf P}^{n_i}(w_{1}^{(i)},\cdots ,w_{n_1+1}^{(i)})$ in the 
$j$-th polynomial $p_j$ ($i=1,2,\cdots ,k\,;\,j=1,2,\cdots ,\ell$).
We propose a formula of the {\kae} potential of the A-model
up to some overall normalization factors
\ba
&&e^{-K}=(-1)^d\int_M\left[
\hat{K}(\lambda)\cdot \overline{\vpi\left(-\frac{\lambda [D_i]}{2\pi
i}\,;\,z_k
\right)}
\cdot \vpi\left(+\frac{\lambda [D_i]}{2\pi i}\,;\,z_k\right)
\right]_{\lambda^d}\times |\vpi_0|^{-2}\,\non
&&=(-1)^d\int_{\cal M}\left[
\hat{K}(\lambda)\cdot \overline{\vpi\left(-\frac{\lambda [D_i]}{2\pi
i}\,;\,z_k\right)}
\cdot \vpi\left(+\frac{\lambda [D_i]}{2\pi i}\,;\,z_k\right)
\right]_{\lambda^d}\cdot [H]\times |\vpi_0|^{-2}\,,\non
&&\vpi (v\,;\,z):=\sum_{n}\frac{a(n+v)}{a(v)}z^{n+v}\,,\,\,\,
z^{n+v}:=\prod_{k=1}^pz^{n_k+v_k}_k\,,\non
&&a(n+v):=
\frac{\dis \prod^{\ell}_{j=1}
\Gamma\!\left(1+\sum_{i=1}^k d_j^{(i)}(n_i+v_i)\right)}
{\dis \prod_{i=1}^k\prod_{j'=1}^{n_i+1}
\Gamma\!\left(1+w_{j'}^{(i)}(n_i+v_i)\right)}\,,\non
&&\hat{K}(\lambda):=
\frac{\dis a\left(+\frac{\lambda [D]}{2\pi i}\right)}
{\dis a\left(-\frac{\lambda [D]}{2\pi i}\right)}
=\exp\left[
+2\sum_{m=1}\frac{\zeta (2m+1)}{2m+1}\cdot \left(\frac{\lambda}{2\pi i}
\right)^{2m+1}\cdot X_{2m+1}\right]\,,\non
&&X_n:=\sum_{i=1}^k\sum_{j'=1}^{n_i +1}(w^{(i)}_{j'}[D_i])^n-
\sum_{j=1}^{\ell}\left(\sum_{i=1}^{k}d_j^{(i)}[D_i]\right)^n
\,,\,\,\,
[H]=\frac{\dis \prod^{\ell}_{j=1}\left(\sum^k_{i=1}d^{(i)}_j[D_i]\right)}
{\dis \prod^k_{i=1}\prod^{n_i+1}_{j'=1}w^{(i)}_{j'}}\,,\non
&&d=-\ell +\sum^{k}_{i=1}n_i\,,\,\,\,\,(\mbox{dimension})\,.\nom
\ea
The $[D_i]$s ($i=1,2,\cdots ,k$) are Poincar\'e duals of divisors ``$D_i$'' 
of the model.
We will rewrite these in order to interpret them 
in the A-model language.
First mirror maps are given as
\ba
&&2\pi i\,t^i=\log z^i+\del_{v_i}\log \hat{\vpi}\,,\non
&&\hat{\vpi}=\sum^{\infty}_{n=0}\frac{a(n+v)}{a(v)}z^n\,.\nom
\ea
Then the normalized $\varpi$ is expressed as
\ba
&&\vpi_{0}^{-1}\cdot \vpi\left(\frac{\lambda [D_i]}{2\pi i}\right)=
\exp \left(\lambda[D_i]\cdot t^i+\sum_{\ell
=2}\lambda^{\ell}x_{\ell}\right)\,,\non
&&x_{\ell}=\frac{1}{(2\pi i)^{\ell}}\frac{1}{\ell !}([D]\cdot
\del)^{\ell}\log \hat{\vpi}\,,\non
&&[D]\cdot\del:=\sum_{i}[D_i]\frac{\del}{\del v_i}\,,\non
&&\hat{K}(\lambda)=\exp \left(\sum_{m=1}\lambda^{2m+1}\hat{k}_{2m+1}
\right)\,,\non
&&\hat{k}_{2m+1}=2\cdot \frac{X_{2m+1}}{2m+1}
\frac{\zeta (2m+1)}{(2\pi i)^{2m+1}}\,.\nom
\ea
By using the above relations, we can express the {\kae} potential 
in a compact form
\ba
&&e^{-K}=(-1)^d\int_{\cal M}S_d(\tilde{y}_1,\tilde{y}_2,\cdots ,\tilde{y}_d)
\cdot [H]\,,\label{find3}\\
&& \tilde{y}_1=\sum_{i} [D_i](t^i-\bar{t}^i)\,,\non
&&\tilde{y}_{2n}=x_{2n}+\bar{x}_{2n}\,,\,\,\,\,(n\geq 1)\,,\non
&&\tilde{y}_{2n+1}=x_{2n+1}-\bar{x}_{2n+1}+
\hat{k}_{2n+1}\,,\,\,\,\,(n\geq 1)\,.\nom
\ea
To confirm the validity of this formula, we restrict ourselves to the 
$3$-fold case and consider the $K$.
First let us consider a general type of Calabi-Yau $3$-fold $M$ with 
{\kae} parameters $t^i$ ($i=1,2,\cdots ,h^{1,1}$).
Its prepotential $F$ is described as
\ba
&&F=-\frac{1}{6}\kei_{ijk}t^it^jt^k+\frac{1}{2}a_{ij}t^it^j+b_it^i
+\frac{1}{2}c+f\,,\non
&&e^{-K}=(t^i-\bar{t}^i)(\del_iF+\bar{\del}_i\bar{F})-2(F-\bar{F})\non
&&\qquad =-\frac{1}{6}\kei_{ijk}(t^i-\bar{t}^i)
(t^j-\bar{t}^j)(t^k-\bar{t}^k)-(a_{ij}-\bar{a}_{ji})t^i\bar{t}^j\non
&&\qquad -(b_i-\bar{b}_i)(t^i-\bar{t}^i)-(c-\bar{c})+
(t^i-\bar{t}^i)(\del_i f+\bar{\del}_i\bar{f})-2(f-\bar{f})\,.\nom
\ea
When we impose a condition that the $e^{-K}$ is invariant under a
constant shift of $t^i\rightarrow t^i +1$ for an arbitrary $i$, 
the term $(a_{ij}-\bar{a}_{ji})t^i\bar{t}^j$ must vanish. 
It means that the ${a_{ij}}$s must be hermitian as components of a matrix
\ba
\bar{a}_{ij}=a_{ji}\,.\nom
\ea
Also the $b_i$s are related to a second Chern class $c_{2}$ of the $M$
\ba
&&b_i=\frac{1}{24}\int_{M}c_{2}\wedge J_i 
=\int_{M}\sqrt{\hat{A}}\wedge J_i\,,\nom
\ea
and they are real numbers. 
On the other hand, the $c$ is a pure imaginary number and is 
associated with a 3rd Chern class
\ba
c=\frac{c_3\zeta (3)}{(2\pi i)^3}\,.\nom
\ea
Collecting all these facts, we can rewrite the $K$
\ba
&&e^{-K}=
-\frac{1}{6}\kei_{ijk}(t^i-\bar{t}^i)
(t^j-\bar{t}^j)(t^k-\bar{t}^k)-2c\non
&&\qquad +(t^i-\bar{t}^i)(\del_i f+\bar{\del}_i\bar{f})-2(f-\bar{f})\,.
\label{3fold}
\ea
The Eq.(\ref{3fold}) is obtained by using a prepotential 
$F$ for a $3$-fold. In contrast,  
we can obtain our result for the formula $K$
without requiring an existence of the prepotential.
Let us write down our result Eq.(\ref{find3}) in the 3 dimensional case.
All we have to do is to evaluate the $S_3$ in this case
\ba
&&S_3=\frac{1}{6}[D_i][D_j][D_k](t-\bar{t})^i
(t-\bar{t})^j(t-\bar{t})^k\non
&&\qquad +[D_i](t-\bar{t})^i(x_2+\bar{x}_2)+(x_3-\bar{x}_3)+\hat{k}_3\,.\nom
\ea
Then the $K$ is obtained
\ba
&&e^{-K}=
-\frac{1}{6}\kei_{ijk}(t-\bar{t})^i
(t-\bar{t})^j(t-\bar{t})^k\non
&&+(t-\bar{t})^i(\del_{t^i}f+\del_{\bar{t}^i}\bar{f})
-2(f-\bar{f})-2\frac{c_3\zeta (3)}{(2\pi i)^3}\,,\label{find4}\\
&&\kei_{ijk}=\int_{\cal M}[D_i][D_j][D_k]\cdot [H]\,,\non
&&\del_{t^i}f=-\frac{1}{2}\frac{1}{(2\pi i)^2}\kei_{ijk}
\frac{\del}{\del v_j}\frac{\del}{\del v_k}\log \hat{\vpi}\,,\non
&&f=\frac{1}{12}\frac{1}{(2\pi i)^3}\kei_{ijk}\frac{\del}{\del v_i}
\frac{\del}{\del v_j}\frac{\del}{\del v_k}\log \hat{\vpi}\,,\non
&&c_3=\frac{1}{3}\left[\sum_i\sum_{j'}(w^{(i)}_{j'})^3\kei_{iii}-
\sum_{i,j,k}\sum_{j'}d^{(i)}_{j'}d^{(j)}_{j'}d^{(k)}_{j'}\kei_{ijk}
\right]\,.\nom
\ea
This formula Eq.(\ref{find4}) 
coincides with that of the $3$-fold case, Eq.(\ref{3fold}). 
But Eq.(\ref{find3}) is not restricted to
this 3 dimensional case. In fact, Eq.(\ref{find3}) is
applicable to smooth 
$d$-folds cases because the periods $\sqrt{\hat{K}}\cdot\varpi$
have appropriate intersection forms
\ba
&&\hat{K}(\lambda)\cdot \overline{\vpi 
\left(-\frac{\lambda [D_i]}{(2\pi i)}\right)}\cdot 
{\vpi \left(+\frac{\lambda [D_i]}{(2\pi i)}\right)}\Biggr|_{\lambda^d}
=\sum^{d}_{n,n'=0}
\bar{\Omega}_n\cdot \Sigma_{n,n'}\cdot \Omega_{n'}\,,\non
&&\Omega_m=
\frac{1}{m!}\frac{1}{(2\pi i)^m}([D]\cdot \del)^m
\left[\sqrt{\hat{K}}\cdot\vpi \right]\,,\,\,\Sigma_{n,n'}=
(-1)^n\delta_{n+n',d}\,.\nom
\ea
For singular $d$-fold case, we might have to modify some parts 
which have information about 
intersection numbers associated with the divisors.

\section{Conclusions and Discussions}
In this article, we develop a method to calculate 
the {\kae} potential in the topological
A-model.
Generally the {\kae} potential $K$ is represented as $\sim (t-\bar{t})^d$ when
no quantum corrections exist.
But it is known that there are loop corrections in the two dimensional $N=2$ 
non-linear sigma models with Calabi-Yau target spaces.
A term $\sim (t-\bar{t})^{d-3}\times 
\dis \frac{c_3\zeta (3)}{(2\pi i)^3} $ in the $K$ reflects 
a perturbative correction at the four loop calculation.
The $c_3$ is the 3rd Chern class of the CY.
In our case, 
the $\hat{K}$ seems to describe loop corrections to the sigma model.
In general, there might be more corrections and they could be controlled 
by one function $\hat{K}$.
It is interesting to give a field theoretical interpretation to these 
$\hat{K}_n$s from the point of view of direct higher loop calculations.

The basis we pick here is a symplectic (or an SO-invariant) 
basis with intersection matrix $\Sigma$.
But it is not integrable but rational basis.
In order to obtain a set of canonical basis, we have to do 
some linear transformation on the $\Omega$. It is needed to discuss 
the D-branes charges or central charges in BPS mass formulae.
We will study these topics in the next paper.

\section*{Acknowledgment}
This work is supported by the Grant-in-Aid for 
Scientific Research from the Ministry of Education, Science and
Culture 10740117.

\newpage
\appendix
\section{Examples of $\hat{K}$}
We write down several concrete examples of the $\hat{K}_n$ 
for the $d$-fold $M$. First the 
generating function $\hat{K}(\rho)$ is defined by using 
Riemann's zeta functions 
\ba
&&\hat{K}(\rho)=\exp\left(2\sum_{m=1}\frac{N-N^{2m+1}}{2m+1}
\zeta (2m+1)\cdot \left(\frac{\rho}{2\pi i}\right)^{2m+1}\right)\non
&&\qquad =1+\sum_{n=1}^d
\hat{K}_n\cdot \left(\frac{\rho}{2\pi i}\right)^n\,,\,\,\,\,\,\,
N=d+2\,.\nom
\ea
The coefficients in the series expansion are represented as some
combinations of Chern classes of the $M$
\ba
&&\hat{K}_{1}=\hat{K}_{2}=0\,,\,\,\,
\hat{K}_{3}=\frac{2}{N}c_3\,\zeta (3)\,,\,\,\hat{K}_{4}=0\,,\non
&&\hat{K}_{5}=\left(\frac{2}{N}c_5-\frac{2}{N^2}c_3c_2\right)\zeta
(5)\,,\,\,\,
\hat{K}_{6}=\frac{2}{N^2}c_3^2 \zeta (3)^2\,,\non
&&\hat{K}_{7}=\left(\frac{2}{N}c_7-\frac{2}{N^2}c_5c_2
-\frac{2}{N^2}c_4c_3+\frac{2}{N^3}c_3c_2^2\right)\zeta (7)\,,\non
&&\hat{K}_{8}=\left(\frac{4}{N^2}c_5c_3 -\frac{4}{N^3}c_3^2c_2\right)
\zeta (3)\zeta (5)\,.\nom
\ea
A finite number of $\hat{K}_n$s ($n\leq d$) appear for the $d$-fold case. 
The $\hat{K}_1$ and $\hat{K}_2$ always vanish and the 
$\hat{K}(\rho)$ is identity for the torus and K3 cases ($N=3,4$
respectively).
We will summarize several examples for lower dimensional cases
\ba
&&\hat{K}(N=3)=
1\,,\,\,\,
\hat{K}(N=4)=
1\,,\non
&&\hat{K}(N=5)=
1 
- 80\,\Mfunction{\zeta}(3)\,{{\Mfunction{\left(\frac{\rho}{2\pi i}\right)}}^3} 
\,,\,\,\,
\hat{K}(N=6)=
1 
- 140\,\Mfunction{\zeta}(3)\,{{\Mfunction{\left(\frac{\rho}{2\pi i}\right)}}^3}\,,\non
&&\hat{K}(N=7)=
1 - 224\,\Mfunction{\zeta}(3)\,{{\Mfunction{\left(\frac{\rho}{2\pi i}\right)}}^3} - 
  6720\,\Mfunction{\zeta}(5)\,{{\Mfunction{\left(\frac{\rho}{2\pi i}\right)}}^5}
\,,\non
&&\hat{K}(N=8)=
1 - 336\,\Mfunction{\zeta}(3)\,{{\Mfunction{\left(\frac{\rho}{2\pi i}\right)}}^3} - 
  13104\,\Mfunction{\zeta}(5)\,{{\Mfunction{\left(\frac{\rho}{2\pi i}\right)}}^5} + 
  56448\,{{\Mfunction{\zeta}(3)}^2}\,{\left(\frac{\rho}{2\pi
        i}\right)^6} 
\,,\non
&&\hat{K}(N=9)=
1 - 480\,\Mfunction{\zeta}(3)\,{{\Mfunction{\left(\frac{\rho}{2\pi i}\right)}}^3} - 
  23616\,\Mfunction{\zeta}(5)\,{{\Mfunction{\left(\frac{\rho}{2\pi
            i}\right)}}^5} \non
&&\qquad + 
  115200\,{{\Mfunction{\zeta}(3)}^2}\,{\left(\frac{\rho}{2\pi i}\right)^6} - 
  1366560\,\Mfunction{\zeta}(7)\,{{\Mfunction{\left(\frac{\rho}{2\pi i}\right)}}^7} 
\,,\non
&&\hat{K}(N=10)=
1 - 660\,\Mfunction{\zeta}(3)\,{{\Mfunction{\left(\frac{\rho}{2\pi i}\right)}}^3} - 
  39996\,\Mfunction{\zeta}(5)\,{{\Mfunction{\left(\frac{\rho}{2\pi
            i}\right)}}^5} \non
&&\qquad + 
  217800\,{{\Mfunction{\zeta}(3)}^2}\,{\left(\frac{\rho}{2\pi i}\right)^6} - 
  2857140\,\Mfunction{\zeta}(7)\,{{\Mfunction{\left(\frac{\rho}{2\pi i}\right)}}^7} + 
  26397360\,\Mfunction{\zeta}(3)\,\Mfunction{\zeta}(5)\,{{\Mfunction{\left(\frac{\rho}{2\pi i}\right)}}^8}\,.\nom
\ea

\end{document}